\documentclass[prd,aps,showpacs,epsf,floats,onecolumn]{revtex4}%
\usepackage{amssymb}
\usepackage{amsfonts}
\usepackage{amsmath}
\usepackage{graphicx}%
\providecommand{\U}[1]{\protect\rule{.1in}{.1in}}

\begin{document}
\title{\textbf{Thermodynamic aspects of information transfer in complex dynamical
systems}}
\author{Carlo Cafaro$^{1}$, Sean Alan Ali$^{2}$, and Adom Giffin$^{3}$}
\affiliation{$^{1}$SUNY Polytechnic Institute, 12203 Albany, New York, USA }
\affiliation{$^{2}$Albany College of Pharmacy and Health Sciences, 12208 Albany, New York, USA}
\affiliation{$^{3}$Clarkson University, 13699 Potsdam, New York, USA}

\begin{abstract}
From the Horowitz-Esposito stochastic thermodynamical description of
information flows in dynamical systems [J. M. Horowitz and M. Esposito, Phys.
Rev. \textbf{X4}, 031015 (2014)], it is known that while the second law of
thermodynamics is satisfied by a joint system, the entropic balance for the
subsystems is adjusted by a term related to the mutual information exchange
rate between the two subsystems. In this article, we present a quantitative
discussion of the conceptual link between the Horowitz-Esposito analysis and
the Liang-Kleeman work on information transfer between dynamical system
components [X. S. Liang and R. Kleeman, Phys. Rev. Lett. \textbf{95}, 244101
(2005)]. In particular, the entropic balance arguments employed in the two
approaches are compared. Notwithstanding all differences between the two
formalisms, our work strengthens the Liang-Kleeman heuristic balance reasoning
by showing its formal analogy with the recent Horowitz-Esposito thermodynamic
balance arguments.

\end{abstract}

\pacs{Thermodynamics (05.70.-a), Entropy (89.70.Cf), Information Theory (89.19.lo),
Conservation Laws (11.30.-j).}
\maketitle

\section{Introduction}

There are several factors that drive the merging of information-theoretic
methods to study the physics of complex dynamical systems. For instance, an
important motivation is provided by the fact that the interacting components
(that is, subsystems) of physical systems generate information at a nonzero
rate \cite{schreiber2000} and exchange information as they influence each
other \cite{horowitz2014}. A quantitative description for the processes of
production, gathering, and exchange of information can be provided by means of
the concept of entropy. In turn, the notions of information and entropy lead
naturally to thermodynamic arguments \cite{parrondo2015}. Indeed, the second
law of thermodynamics is a fundamental guiding principle that imposes
fundamental limits to the amount of information that can be gathered
\cite{lloyd1989} and/or exchanged \cite{horowitz2015} between interacting
components of a physical system.

The interest in describing and, to a certain extent, understanding the
dynamics of information transport in complex systems is justified by the
important role that information transfer analysis has in detecting asymmetry
in the interaction of subsystems \cite{schreiber2000}, in predicting the
weather \cite{liang2005}, in controlling a system \cite{lloyd2000, lloyd2004},
in inferring causal structures \cite{cafaro2014a, cafaro2015b}, and, from a
more conceptual standpoint, in investigating the thermodynamics of Maxwell's
demon \cite{vedral2009, cafaro2013, cafaro2014}.

A convenient way to understand the relation between information and
thermodynamics is by investigating the manner in which information flow
affects the thermodynamics of a system. Two important quantities in
information theory and thermodynamics are given by the rate of mutual
information \cite{cover} and the thermodynamic entropy production
\cite{schnakenberg, lebowitz, seifert12}, respectively. In particular, entropy
production is a central physical observable in stochastic thermodynamics
\cite{seifert12, seifert08}, a framework to study systems far from
equilibrium. Entropic rates of information-theoretic nature, instead, are
central quantities in information theory. Therefore, it seems reasonable to
expect that the investigation of the relation between thermodynamic and
information-theoretic entropic rates can pave the way to a better
understanding of the link between information and thermodynamics. Indeed, this
reasonable expectation has been already explicitly tested in a quantitative
manner in several works in the literature \cite{barato13, barato13D,
barato14C}. To the best of our knowledge, one of the first attempts to
investigate the thermodynamics aspects of information flow with special regard
to the inference of causal structures was presented in \cite{janzing2009},
where a version of the second law of thermodynamics was used to characterize
the thermodynamic cost for information flow in a system defined by a pair of
Brownian particles, each coupled to a thermal bath at different temperatures.
A similar type of hybrid information-theoretic and statistical physics
approach, including reasonings based upon the second law, to the study of
information flow in interacting systems has been recently discussed by
Hartich-Barato-Seifert and Horowitz-Esposito in Refs. \cite{hartich14} and
\cite{horowitz2014}, respectively. More specifically, in \cite{janzing2009}
entropic rate is defined as the time derivative of the time-delayed mutual
information and is used to characterize the information flow between two
Brownian particles coupled to different heat baths, as mentioned earlier. In
\cite{hartich14}, entropic rate is defined as the entropy reduction rate of a
subsystem due to its coupling to the other subsystem composing the bipartite
system and is used to show that Maxwell's demon can be realized in terms of a
bipartite Markov process. Finally, in \cite{horowitz2014} entropic rate is
defined as the time derivative of the mutual information and is used to
investigate the thermodynamics of information flow between two interacting
subsystems of a Markovian bipartite system.

The notion of entropy rate in information transport mechanisms is not uniquely
defined in the literature but it is not the scope of this article to provide a
review of such diverse definitions. Instead, inspired by the works in
\cite{janzing2009, hartich14, horowitz2014}, we focus on two main points in
this article. First, we point out the conceptual link between the
Liang-Kleeman notion of information transfer \cite{liang2005} and the
Horowitz-Esposito notion of information flow \cite{horowitz2014}. This allows
us to avoid some heuristic arguments in the Liang-Kleeman approach and replace
them with thermodynamic reasoning. This first point sets the Liang-Kleeman
analysis on more foundational grounds and, in addition, exhibits its
similarity with the Horowitz-Esposito thermodynamic analysis of information
flow. Second, we compare the foundationally strengthened Liang-Kleeman notion
of information transfer with that provided by Schreiber, after having briefly
discussed the physical \cite{ito2013} and thermodynamical
\cite{prokopenko2013} interpretations of the latter. This second point
improves our understanding of the differences between the two notions from a
more fundamental point of view \cite{janzing2009}.

The layout of this article is as follows. In Sec. II, we reexamine the
entropic aspects of both the Liang-Kleeman and the Horowitz-Esposito
approaches. In Sec. III, we strengthen the Liang-Kleeman heuristic entropic
balance arguments by emphasizing the formal analogies between the two
approaches. We also compare such thermodynamically strengthened information
transfer measure with the thermodynamic interpretation of Schreiber's transfer
entropy. Finally, in Sec. IV we present our conclusive remarks.

\section{Conservation of probability}

The transport of a conserved quantity is described by a continuity equation
whose differential form is given by \cite{huang1987},%
\begin{equation}
\frac{\partial\rho}{\partial t}+\vec{\nabla}\cdot\vec{j}=0\text{,}
\label{continuity}%
\end{equation}
where $\rho$ denotes the volume density of the conserved quantity and $\vec
{j}$ is the ordinary vector flux of the conserved quantity. In classical
electrodynamics, for instance, the electric charge is the conserved quantity,
$\rho$ is the electric charge density, and $\vec{j}$ is the electric current
density \cite{cafaro-ali}. A similar line of reasoning appears in fluid
dynamics, thermodynamics, and quantum theory where the electric charge is
replaced with mass, heat, and probability distributions, respectively. A
continuity equation associated to the conservation of probability is the
fundamental starting point of discussion in both the Liang-Kleeman and the
Horowitz-Esposito approaches to information transfer and information flow.

\subsection{The Liang-Kleeman approach: Heuristic arguments}

The Liang-Kleeman approach can be applied to both deterministic and stochastic
systems of arbitrary finite dimensionality, either discrete or continuous. For
a recent comprehensive review of the Liang-Kleeman approach, we refer to
\cite{liang2013}. For the latest developments on this approach, we refer
instead to \cite{liang2015}. In this article, our discussion is based upon the
working hypotheses of the original work presented in \cite{liang2005} which
contains the essence of the philosophy of the Liang-Kleeman approach.
Specifically, we shall be considering a two-dimensional continuous and
deterministic autonomous system with \emph{a priori} knowledge of the
dynamics,%
\begin{equation}
\frac{d\vec{x}}{dt}=\vec{F}\left(  \vec{x}\right)  \text{,} \label{det}%
\end{equation}
where $\vec{x}=\left(  x_{1}\text{, }x_{2}\right)  $ belongs to the\textbf{
}state space $\Omega=\Omega_{1}\times\Omega_{2}$ and $\vec{F}=\left(
F_{1}\text{, }F_{2}\right)  $ with $F_{i}=F_{i}\left(  x_{1}\text{, }%
x_{2}\right)  $ for any $i=1$, $2$ is the (known) flow vector. The sample
values $\left(  x_{1}\text{, }x_{2}\right)  $ are associated to a stochastic
process $\vec{X}=\left(  X_{1}\text{, }X_{2}\right)  $ with (known) joint
probability density distribution at time $t$ given by $\rho=\rho\left(
x_{1}\text{, }x_{2}\text{, }t\right)  $. The vector flux $\vec{j}$ equals
$\rho\vec{F}$ and the continuity equation associated with Eq. (\ref{det})
becomes,%
\begin{equation}
\frac{\partial\rho}{\partial t}+\vec{\nabla}\cdot\left(  \rho\vec{F}\right)
=0\text{.} \label{de}%
\end{equation}
Assuming that $\rho$ vanishes at the boundaries, after some suitable algebraic
manipulations of Eq. (\ref{de}), it is found that the temporal rate of change
of the joint entropy of $X_{1}$ and $X_{2}$,%
\begin{equation}
H\left(  t\right)  \overset{\text{def}}{=}-\underset{\Omega}{\int\int}\rho
\log\rho dx_{1}dx_{2}\text{,} \label{shannone}%
\end{equation}
satisfies the relation \cite{liang2005},%
\begin{equation}
\frac{dH}{dt}=\mathbb{E}\left(  \vec{\nabla}\cdot\vec{F}\right)  \text{.}
\label{teq}%
\end{equation}
The base of the logarithm in Eq. (\ref{shannone}) determines the units used
for measuring information: bits and nats for base-$2$ and base-$e$,
respectively. Equation (\ref{teq}) states the $dH/dt$ equals the expectation
value of the divergence of the flow vector $\vec{F}$,
\begin{equation}
\mathbb{E}\left(  \vec{\nabla}\cdot\vec{F}\right)  \overset{\text{def}}%
{=}\underset{\Omega}{\int\int}\rho\left(  \vec{\nabla}\cdot\vec{F}\right)
dx_{1}dx_{2}\text{.}%
\end{equation}
To describe the decomposition of the various mechanisms responsible for the
joint and individual temporal rates of changes of entropies of $X_{1}$,
$X_{2}$, and $\left(  X_{1}\text{, }X_{2}\right)  $ in terms of information
transfers, Liang and Kleeman employ a very clever heuristic argument. First,
following the very same line of reasoning and working hypotheses used to
obtain Eq. (\ref{teq}), they compute both $dH_{1}/dt$ and $dH_{2}/dt$ where
$H_{i}$ denotes the entropy of $X_{i}$ defined in terms of $\rho_{i}$ obtained
from $\rho$ after marginalizing over the degree of freedom $j$ with $j\neq i$.
Second, they observe that if $X_{2}$ is frozen and $X_{1}$ evolves on its own,
its entropic rate of change would be equal to $\mathbb{E}\left(  \partial
F_{1}/\partial x_{1}\right)  $. In the presence of interacting processes
$X_{1}$ and $X_{2}$ they find that,%
\begin{equation}
\frac{dH_{1}}{dt}\neq\mathbb{E}\left(  \frac{\partial F_{1}}{\partial x_{1}%
}\right)  \overset{\text{def}}{=}\frac{dH_{1}^{\ast}}{dt}\text{.} \label{teq0}%
\end{equation}
Therefore, they arrive at the conclusion that the difference between
$dH_{1}/dt$ and $\mathbb{E}\left(  \partial F_{1}/\partial x_{1}\right)  $
must equal the rate of entropy transfer from $X_{2}$ to $X_{1}$, and define
transfer entropy as%
\begin{equation}
T_{2\rightarrow1}\overset{\text{def}}{=}\frac{dH_{1}}{dt}-\frac{dH_{1}^{\ast}%
}{dt}=-\underset{\Omega}{\int\int}\rho_{2\left\vert 1\right.  }\left(
x_{2}\left\vert x_{1}\right.  \right)  \frac{\partial\left(  \rho_{1}%
F_{1}\right)  }{\partial x_{1}}dx_{1}dx_{2}\text{,} \label{teq1}%
\end{equation}
where $\rho_{2\left\vert 1\right.  }\left(  x_{2}\left\vert x_{1}\right.
\right)  \overset{\text{def}}{=}\rho\left(  x_{1}\text{, }x_{2}\text{,
}t\right)  /\rho_{1}\left(  x_{1}\text{, }t\right)  $. Observe that combining
Eqs. (\ref{teq}), (\ref{teq0}), and (\ref{teq1}), we obtain%
\begin{equation}
\frac{dH_{1}}{dt}+\frac{dH_{2}}{dt}=\frac{dH}{dt}+T_{1\rightarrow
2}+T_{2\rightarrow1}\text{.} \label{IMPO1}%
\end{equation}
At this juncture we make the following remarks. First, when a flow is
additive, matter has to be lost in one component of the system in order for
another component to receive it. However, information does not have to be lost
in one component in order for another component to receive it. A key
difference between flow of matter and flow of information is encoded into the
asymmetry of the latter. Specifically, we note that if $F_{1}=F_{1}\left(
x_{1}\right)  $ and does not depend on $x_{2}$, we have $T_{2\rightarrow1}=0$
and there is no information transfer from $X_{2}$ to $X_{1}$. As a result of
the asymmetric nature of information transfer however, this does not
necessarily imply that $T_{1\rightarrow2}=0$. We point out that the idea of
frozen variables also appears in the framework of bipartite networks
\cite{barato13} where one considers Markov processes with states that are
specified by two variables. In a transition between states, only one of the
variables can change while the other one is kept fixed. A few recent and
interesting applications of such bipartite network formalism applied to
stochastic modeling for cellular and biochemical sensing appear in Refs.
\cite{barato13D} and \cite{barato14C}, respectively. Second, the physical
consistency of the Liang-Kleeman approach with the Schreiber work
\cite{schreiber2000} (see also Section III) can be explained as follows. In
the former approach, a direction of the phase space is frozen in order to
extract information transfer. In the latter, the transition probabilities are
essentially obtained from joint probabilities by fixing one component of the
joint system. Third, a conserved information-theoretic charge can be defined
as,%
\begin{equation}
Q\left(  t\right)  \overset{\text{def}}{=}H\left(  t\right)  -\int^{t}E\left(
\vec{\nabla}\cdot\vec{F}\right)  dt^{\prime}\text{.}%
\end{equation}
In general, $Q\left(  t\right)  $ is not an additive quantity and it becomes
so if one assumes $\rho=\rho_{1}\rho_{2}$, $F_{1}=F_{1}\left(  x_{1}\right)
$, and $F_{2}=F_{2}\left(  x_{2}\right)  $. In other words, the conserved
charge $Q\left(  t\right)  $ is additive and equals $Q_{1}\left(  t\right)
+Q_{2}\left(  t\right)  $ if and only if there is no information flow between
the components $X_{1}$ and $X_{2}$ of the joint system $\left(  X_{1}\text{,
}X_{2}\right)  $. We therefore conclude that the presence of information
transfers between the components of a complex system is incompatible with the
additivity of a conserved information-theoretic charge.

\subsection{The Horowitz-Esposito approach: Thermodynamic arguments}

The second law of thermodynamics states that the entropy of an isolated
macroscopic system cannot decrease in time \cite{fermi}. If fluctuation
effects are also taken into account, it is possible to provide a new version
of the second law and, more generally, a new thermodynamic description for
small systems. Such a framework is known as stochastic thermodynamics
\cite{seifert12, seifert08, broeck}. In Ref. \cite{hartich14}, being within
the framework of stochastic thermodynamics, Hartich-Barato-Seifert have
investigated various second-law-like inequalities valid for bipartite systems.
In particular, they have provided an especially clear thermodynamical
interpretation of Maxwell's demon. Specifically, they have considered a
bipartite system ($xy$-system) where Maxwell's demon is one of the subsystems
($y$-system) that affects the other subsystem ($x$-system) used to extract
work from a heat bath. Here we focus on the stochastic thermodynamical
formalism used in \cite{horowitz2014} by Horowitz and Esposito to characterize
how information flow influences the thermodynamics of a system described by a
probability distribution that evolves according to a Markovian master equation
\cite{esposito2010a} (or, more specially, a Fokker-Planck equation
\cite{esposito2010b}).

In what follows, we limit our discussion to bipartite thermodynamics and
information flow \cite{horowitz2014}. For a very recent analysis extended to
multipartite flow, we refer to \cite{horowitz2015}. Specifically, consider a
Markovian system $\left(  X\text{, }Y\right)  $ described by a joint
probability distribution $p\left(  x\text{, }y\right)  $ that evolves
according to a master equation,%
\begin{equation}
\frac{dp\left(  x\text{, }y\right)  }{dt}-\sum_{x^{\prime}\text{, }y^{\prime}%
}\left[  W_{x\text{, }x^{\prime}}^{y\text{, }y^{\prime}}p\left(  x^{\prime
}\text{, }y^{\prime}\right)  -W_{x^{\prime}\text{, }x}^{y^{\prime}\text{, }%
y}p\left(  x\text{, }y\right)  \right]  =0\text{,} \label{me}%
\end{equation}
where $W_{x\text{, }x^{\prime}}^{y\text{, }y^{\prime}}$ is the the transition
rate (matrix) at which the system $\left(  X\text{, }Y\right)  $ jumps from
$\left(  x^{\prime}\text{, }y^{\prime}\right)  $ to $\left(  x\text{,
}y\right)  $. An important working hypothesis in \cite{horowitz2014} is the
so-called bipartite assumption according to which the fluctuations (noises) in
each subsystem $X$ and $Y$ are independent.\ Although these fluctuations are
not required to be generated by distinct thermodynamics reservoirs, Horowitz
and Esposito employ this additional assumption for the sake of simplicity of
exposition of their original argument. Equation (\ref{me}) can be recast as a
continuity equation as in Eq. (\ref{continuity}) since probability is
conserved. To do so, the current (probability flux) $J$ flowing from $\left(
x^{\prime}\text{, }y^{\prime}\right)  $ to $\left(  x\text{, }y\right)  $ is
defined as,%
\begin{equation}
J_{x\text{, }x^{\prime}}^{y\text{, }y^{\prime}}\overset{\text{def}}%
{=}W_{x\text{, }x^{\prime}}^{y\text{, }y^{\prime}}p\left(  x^{\prime}\text{,
}y^{\prime}\right)  -W_{x^{\prime}\text{, }x}^{y^{\prime}\text{, }y}p\left(
x\text{, }y\right)  \text{.} \label{j}%
\end{equation}
Combining Eqs. (\ref{me}) and (\ref{j}), the continuity equation becomes%
\begin{equation}
\frac{dp\left(  x\text{, }y\right)  }{dt}-\sum_{x^{\prime}\text{, }y^{\prime}%
}J_{x\text{, }x^{\prime}}^{y\text{, }y^{\prime}}=0\text{.} \label{de2}%
\end{equation}
A key point in the Horowitz-Esposito approach that plays an essential role in
our comparison with the Liang-Kleeman work is the exploitation of the
bipartite structure in order to split the current $J_{x\text{, }x^{\prime}%
}^{y\text{, }y^{\prime}}$ into two separate flows, $J_{x\text{, }x^{\prime}%
}^{y}$ (the current from $x^{\prime}$ to $x$ along $y$) and $J_{x}^{y\text{,
}y^{\prime}}$ (the current from $y^{\prime}$ to $y$ along $x$), in such a
manner that%
\begin{equation}
\sum_{x^{\prime}\text{, }y^{\prime}}J_{x\text{, }x^{\prime}}^{y\text{,
}y^{\prime}}\overset{\text{def}}{=}\sum_{x^{\prime}}J_{x\text{, }x^{\prime}%
}^{y}+\sum_{y^{\prime}}J_{x\text{ }}^{y\text{, }y^{\prime}}\text{.}
\label{flow}%
\end{equation}
Equation (\ref{flow}) implies that the flow in the joint $\left(  X\text{,
}Y\right)  $-direction is the sum of two separate flows, one in the
$X$-direction and one in the $Y$-direction. The reasoning of Horowitz and
Esposito goes as follows. They are interested in describing how the
information flow between the subsystems $X$ and $Y$ is linked to the
thermodynamics of the joint system $\left(  X\text{, }Y\right)  $. First, they
observe that the second law of thermodynamics applied to $\left(  X\text{,
}Y\right)  $ requires that \cite{esposito2010a},%
\begin{equation}
\dot{S}_{i}=d_{t}S^{XY}+\dot{S}_{r}\geq0\text{,} \label{second}%
\end{equation}
where $\dot{S}_{i}$ is the irreversible entropy production rate, $\dot{S}_{r}$
is the entropy rate flowing to the environment, and $d_{t}S^{XY}$ denotes the
time derivative of the joint (Shannon) entropy of $\left(  X\text{, }Y\right)
$. Second, they note that Eq. (\ref{second}) does not explain the manner in
which information flows between $X$ and $Y$. To achieve this goal, they point
out that $\dot{S}_{i}$, $d_{t}S^{XY}$, and $\dot{S}_{r}$ are linear
functionals of the currents, that is, they are flows. Then, in analogy to the
splitting of the current as in Eq. (\ref{flow}), they decompose any current
functional $\mathcal{A}\left[  J\right]  $ as,%
\begin{equation}
\mathcal{A}\left[  J\right]  =\mathcal{A}^{X}+\mathcal{A}^{Y}\text{.}%
\end{equation}
The quantities $\mathcal{A}^{X}$ and $\mathcal{A}^{Y}$ denote the variations
in the $X$-direction and $Y$-direction of $\mathcal{A}\left[  J\right]  $,
respectively. In particular, we find%
\begin{equation}
\dot{S}_{i}=\dot{S}_{i}^{X}+\dot{S}_{i}^{Y}\text{, and }\dot{S}_{r}=\dot
{S}_{r}^{X}+\dot{S}_{r}^{Y}\text{.} \label{dec}%
\end{equation}
Using Eq. (\ref{dec}), recalling the information-theoretic relation
$S^{XY}=S^{X}+S^{Y}-I$ between joint entropy $S^{XY}$ and mutual information
$I$ \cite{cover},%
\begin{equation}
I\overset{\text{def}}{=}\sum_{x\text{, }y}p\left(  x\text{, }y\right)
\log\left[  \frac{p\left(  x\text{, }y\right)  }{p\left(  x\right)  p\left(
y\right)  }\right]  \text{,} \label{ii}%
\end{equation}
and defining the so-called information flow as \cite{horowitz2014},%
\begin{equation}
d_{t}I\overset{\text{def}}{=}\dot{I}^{X}+\dot{I}^{Y}\text{,} \label{dif1}%
\end{equation}
it follows after some simple algebra that Eq. (\ref{second}) can be decomposed
into two equations,%
\begin{equation}
\dot{S}_{i}^{X}=d_{t}S^{X}+\dot{S}_{r}^{X}-\dot{I}^{X}\text{ and, }\dot{S}%
_{i}^{Y}=d_{t}S^{Y}+\dot{S}_{r}^{Y}-\dot{I}^{Y}\text{,} \label{decio}%
\end{equation}
respectively. Observe that combining Eqs. (\ref{second}), (\ref{dec}), and
(\ref{decio}), we obtain%
\begin{equation}
d_{t}S^{X}+d_{t}S^{Y}=d_{t}S^{XY}+\dot{I}^{X}+\dot{I}^{Y}\text{.}
\label{IMPO2}%
\end{equation}
The two relations in Eq. (\ref{decio}) and especially the reasoning underlying
their derivation represent the main result provided by Horowitz and Esposito
that we are concerned with in this article.

\section{A thermodynamic comparison}

In 2000, Schreiber formulated the current broadly accepted
information-theoretic notion of transfer entropy \cite{schreiber2000}.
Transfer entropy is a measure of time-asymmetric information transfer between
jointly distributed random vectors associated with time series observations.
For an interesting outline of general differences between transfer entropy and
information flow defined via a time-shifted mutual information, we refer to
\cite{janzing2009}. For relatively recent thermodynamic interpretations of
transfer entropy, we refer to \cite{ito2013, prokopenko2013}.

In what follows, we compare the thermodynamic interpretation of Schreiber's
transfer entropy as presented in \cite{ito2013, prokopenko2013} with the
Liang-Kleeman measure of information transfer after having emphasized its
formal analogy with the Horowitz-Esposito thermodynamic formalism.

\subsection{The Liang-Kleeman approach: Thermodynamic arguments}

Recall that the mutual information $I$ between $X_{1}$ and $X_{2}$ is formally
defined as \cite{cover},
\begin{equation}
I\left(  t\right)  \overset{\text{def}}{=}\underset{\Omega}{\int\int}\rho
\log\left(  \frac{\rho}{\rho_{1}\rho_{2}}\right)  dx_{1}dx_{2}\text{,}
\label{MI}%
\end{equation}
and equals,%
\begin{equation}
I\left(  t\right)  =H_{1}\left(  t\right)  +H_{2}\left(  t\right)  -H\left(
t\right)  \text{.} \label{above}%
\end{equation}
Differentiating both sides of Eq. (\ref{above}) with respect to time, we
obtain
\begin{equation}
d_{t}I=d_{t}H_{1}+d_{t}H_{2}-d_{t}H\text{.} \label{dif2}%
\end{equation}
By combining Eq. (\ref{IMPO1}) and (\ref{dif2}), we find%
\begin{equation}
d_{t}I=T_{1\rightarrow2}+T_{2\rightarrow1}\text{,} \label{dif22}%
\end{equation}
where $I$ is defined in Eq. (\ref{MI}) and $T_{1\rightarrow2}$ in Eq.
(\ref{teq1}). After having reexamined both the Horowitz-Esposito and the
Liang-Kleeman approaches, the similarities between the two approaches become
almost self-evident. Specifically, Eqs. (\ref{de}), (\ref{IMPO1}), and
(\ref{dif22}) within the Liang-Kleeman approach correspond to Eqs.
(\ref{de2}), (\ref{IMPO2}), and (\ref{dif1}) within the Horowitz-Esposito
approach, respectively. Due to these conceptual and mathematical analogies,
the heuristic entropic balance arguments in \cite{liang2005} are automatically
supported by the thermodynamically grounded entropic balance arguments as
presented in \cite{horowitz2014}. One of our main contributions in this
article is pointing out these hidden connections after a critical
reconsideration of both approaches. We emphasize that the structure of Eq.
(\ref{dif22}) is similar to that of Eq. (B.5) in \cite{hartich14}: the time
derivative of mutual information in Eq. (\ref{ii}) and the Liang-Kleeman
definition of transfer entropy in Eq. (\ref{teq1}) are replaced in
\cite{hartich14} by the continuous time rate of mutual information and the
continuous time version of Schreiber's transfer entropy, respectively.

\subsection{The Schreiber approach: Thermodynamic arguments}

The problem addressed by Schreiber was the following: given time series
observations associated with interacting subsystems of a complex system,
quantify how information produced by an individual subsystem is exchanged with
another subsystem. To address this problem, Schreiber proposed an
information-theoretic measure containing both dynamical and directional
information. To incorporate aspects of the dynamics of information transport
into the measure, Schreiber considered transition rather than static
probabilities. Subsystems are assumed to be modeled in terms of stationary
Markov processes. A Markov process $I$ is of order $k$ if the conditional
probability to find $I$ in the state $i_{n+1}$ does not depend on the state
$i_{n-k}$,%
\begin{equation}
p\left(  i_{n+1}\left\vert i_{n}\text{,..., }i_{n-k+1}\right.  \right)
=p\left(  i_{n+1}\left\vert i_{n}\text{,..., }i_{n-k}\right.  \right)
\text{,}%
\end{equation}
where for notational simplicity we define $i_{n}^{\left(  k\right)  }%
\overset{\text{def}}{=}\left(  i_{n}\text{,..., }i_{n-k+1}\right)  $. Ignoring
static correlations due to common history, transfer entropy is able to detect
the directed exchange of information between two subsystems by measuring the
deviation from the generalized Markov property,%
\begin{equation}
p\left(  i_{n+1}\left\vert i_{n}^{\left(  k\right)  }\right.  \right)
=p\left(  i_{n+1}\left\vert i_{n}^{\left(  k\right)  }\text{, }j_{n}^{\left(
l\right)  }\right.  \right)  \text{.} \label{MP}%
\end{equation}
In the absence of information transfer from $J$ to $I$, the state of $J$ has
no influence on the transition probabilities on system $I$. Transfer entropy
quantifies the departure from the assumption in Eq. (\ref{MP}) and is defined
as%
\begin{equation}
T_{J\rightarrow I}\overset{\text{def}}{=}\sum_{i_{n+1}\text{, }i_{n}^{\left(
k\right)  }\text{, }j_{n}^{\left(  l\right)  }}p\left(  i_{n+1}\text{, }%
i_{n}^{\left(  k\right)  }\text{, }j_{n}^{\left(  l\right)  }\right)
\log\left[  \frac{p\left(  i_{n+1}\left\vert i_{n}^{\left(  k\right)  }\text{,
}j_{n}^{\left(  l\right)  }\right.  \right)  }{p\left(  i_{n+1}\left\vert
i_{n}^{\left(  k\right)  }\right.  \right)  }\right]  \text{.} \label{TE}%
\end{equation}
Noting that $p\left(  i\text{, }j\text{, }k\right)  =p\left(  i\left\vert
j\text{, }k\right.  \right)  p\left(  j\text{, }k\right)  =p\left(  i\text{,
}j\left\vert k\right.  \right)  p\left(  k\right)  $, after some simple
algebraic manipulations, it is found that the transfer entropy in Eq.
(\ref{TE}) can be rewritten as,%
\begin{equation}
T_{J\rightarrow I}=\Delta s_{I\left\vert J\right.  }-\Delta s_{I}\text{,}
\label{TE-termo}%
\end{equation}
where the entropy rates $\Delta s_{I\left\vert J\right.  }$ and $\Delta s_{I}$
are given by,%
\begin{equation}
\Delta s_{I\left\vert J\right.  }\overset{\text{def}}{=}\sum_{i_{n+1}\text{,
}i_{n}^{\left(  k\right)  }\text{, }j_{n}^{\left(  l\right)  }}p\left(
i_{n+1}\text{, }i_{n}^{\left(  k\right)  }\text{, }j_{n}^{\left(  l\right)
}\right)  \left[  \log p\left(  i_{n+1}\text{, }i_{n}^{\left(  k\right)
}\left\vert j_{n}^{\left(  l\right)  }\right.  \right)  -\log p\left(  \text{
}i_{n}^{\left(  k\right)  }\left\vert j_{n}^{\left(  l\right)  }\right.
\right)  \right]  \text{,}%
\end{equation}
and%
\begin{equation}
\Delta s_{I}\overset{\text{def}}{=}\sum_{i_{n+1}\text{, }i_{n}^{\left(
k\right)  }}p\left(  i_{n+1}\text{, }i_{n}^{\left(  k\right)  }\right)
\left[  \log p\left(  i_{n+1}\text{, }i_{n}^{\left(  k\right)  }\right)  -\log
p\left(  \text{ }i_{n}^{\left(  k\right)  }\right)  \right]  \text{,}%
\end{equation}
respectively. From Eq. (\ref{TE-termo}), one concludes that the transfer
entropy from the process $J$ to the process $I$ equals the difference between
the entropy rate in $I$ (under the condition $J$)\ and the entropy rate in
$I$. This physical interpretation of transfer entropy was originally proposed
in \cite{ito2013}. In addition, in Ref. \cite{ito2013} Ito and Sagawa employed
the notion of transfer entropy as defined by Schreiber in their generalized
version of the second law of thermodynamics. They showed that the entropy
production in a single system $J$ interacting with multiple other systems in
$\mathcal{K}=\left\{  K_{1}\text{,..., }K_{N}\right\}  $ is bounded by the
information flow between these systems. Specifically, modeling the interacting
systems by means of a Bayesian network \cite{lloyd1991}, they showed that%
\begin{equation}
\left\langle \Delta I\right\rangle -\left\langle \sigma\right\rangle \leq
\sum_{l=1}^{N}T_{J\rightarrow K_{l}}\text{,}%
\end{equation}
where $\left\langle \Delta I\right\rangle \overset{\text{def}}{=}\left\langle
I_{\text{fin}}\right\rangle -\left\langle I_{\text{ini}}\right\rangle $ is the
difference between the ensemble averages of the final and initial correlations
between $J$ and the other subsystems in $\mathcal{K}$ (that is, the mutual
information exchanged between the systems $J$ and $\mathcal{K}$),
$\left\langle \sigma\right\rangle $ is the ensemble average of the entropy
production $\sigma$ of subsystem $J$, and $T_{J\rightarrow K_{l}}$ is the
transfer entropy that characterizes the information transfer from $J$ to
$K_{l}$. For further technical details, we refer to \cite{ito2013}. We also
remark that, inspired by the fluctuation relation for causal networks obtained
in \cite{ito2013} and using the concept of transfer entropy in the continuos
time limit \cite{barato13}, Barato-Hartich-Seifert have recently presented a
similar fluctuation relation for a stochastic trajectory of a bipartite system
\cite{hartich14}.

In what follows, we shall focus on the thermodynamic interpretation of
Schreiber's transfer entropy as proposed by Prokopenko \textit{et}
\textit{al}. in Ref. \cite{prokopenko2013}. The Prokopenko et\textit{ al}.
approach can be synthesized as follows. For the sake of simplicity, consider a
joint system $\left(  X\text{, }Y\right)  $. For each subsystem, the variation
of entropy $\Delta S$ equals the sum of the entropy change caused by
interactions with the surrounding $\Delta S_{\text{ext}}$ and the internal
entropy production $\sigma$ inside the system itself. Therefore,%
\begin{equation}
\Delta S=\sigma+\Delta S_{\text{ext}}\text{.} \label{ww}%
\end{equation}
Two assumptions are employed. First, the conditional probability $p\left(
x_{n+1}\left\vert x_{n}\right.  \right)  $ is related to the transition
probability of the system's reversible state change. Second, the conditional
probability $p\left(  x_{n+1}\left\vert x_{n}\text{, }y_{n}\right.  \right)  $
is related to the transition probability of the system's possibly irreversible
internal state change, due to the interactions with the external surroundings.
Furthermore, assuming a Boltzmann entropic expression for the Shannon entropy,
Prokopenko \textit{et al}. arrive at the proportionality between local
transfer entropy $t_{Y\rightarrow X}$ and external entropy production $\Delta
S_{\text{ext}}$ in the additional working hypothesis of small fluctuations
near equilibrium \cite{prokopenko2013},%
\begin{equation}
t_{Y\rightarrow X}\left(  n+1\right)  \propto-\Delta S_{\text{ext}}%
^{X}\text{,} \label{proko}%
\end{equation}
where%
\begin{equation}
t_{Y\rightarrow X}\left(  n+1\right)  \overset{\text{def}}{=}\log\left[
\frac{p\left(  x_{n+1}\left\vert x_{n}\text{, }y_{n}\right.  \right)
}{p\left(  x_{n+1}\left\vert x_{n}\right.  \right)  }\right]  \text{,}%
\end{equation}
and transfer entropy $T_{Y\rightarrow X}$ in Eq. (\ref{TE-termo}) is the
expectation value $\left\langle t_{Y\rightarrow X}\left(  n+1\right)
\right\rangle $ of $t_{Y\rightarrow X}$ at each time step $n$. Considering Eq.
(\ref{ww}) for both subsystems, using Eq. (\ref{proko}) and setting
Boltzmann's constant $k_{B}$ equal to one (the proportionality factor in Eq.
(\ref{proko}) is given by $k_{B}\ln2$ when entropy is given in units of nats),
it is determined that%
\begin{equation}
\sigma^{X}+\sigma^{Y}=\Delta S+\left[  t_{Y\rightarrow X}\left(  n+1\right)
+t_{Y\rightarrow X}\left(  n+1\right)  \right]  \text{,} \label{IMPO3}%
\end{equation}
where $\Delta S$ in Eq. (\ref{IMPO3}) equals $\Delta S^{\left(  X\text{,
}Y\right)  }$. Equation (\ref{IMPO3}) is the analog of Eqs. (\ref{IMPO1}) and
(\ref{IMPO2}) within the Liang-Kleeman and Horowitz-Esposito approaches,
respectively. We remark that while these analogies are quite illuminating from
a thermodynamic standpoint, they must be taken \emph{cum grano salis}. For
instance, unlike the information flow $d_{t}I=\dot{I}^{X}+\dot{I}^{Y}$ in Eq.
(\ref{dif1}), transfer entropy as defined by Schreiber in Eq. (\ref{TE}) is
not a flow since it does not add up additively to time derivative of any
global quantity. Furthermore, unlike the information flow, transfer entropy is
always non-negative. For a recent and detailed information-theoretic analysis
on the use of transfer entropy rate and information flow to bound the
extracted work during a thermodynamic process with feedback, we refer to
\cite{horowitznjp}.

Despite formal mathematical and trivial notational differences, we have
provided here a novel quantitative reexamination leading to the essential
thermodynamical and physical similarities among the Liang-Kleeman
\cite{liang2005}, the Horowitz-Esposito \cite{horowitz2014}, and the Schreiber
\cite{schreiber2000} works.

\section{Conclusions}

The stochastic thermodynamical approach to information flows in dynamical
systems as presented by Horowitz and Esposito \cite{horowitz2014} explains
that while the second law of thermodynamics is satisfied by the joint system,
the entropic balance for the subsystems is adjusted by a term related to the
mutual information rate between the two subsystems. We have conveniently
recast the Horowitz-Esposito formalism in Eq. (\ref{IMPO2}) as,%
\begin{equation}
d_{t}S^{X}+d_{t}S^{Y}=d_{t}S^{XY}+\left[  \dot{I}^{X}+\dot{I}^{Y}\right]
\text{.} \label{A}%
\end{equation}
In this article, starting from the conservation of probability and the
continuity equation (see Eqs. (\ref{continuity}), (\ref{de}), and
(\ref{de2})), we presented a quantitative discussion on the physical link
between the Horowitz-Esposito analysis \cite{horowitz2014} and the
Liang-Kleeman work on information transfer between dynamical system components
\cite{liang2005}. In particular, the entropic balance arguments employed in
the two approaches were compared. Notwithstanding all differences between the
two formalisms, our work strengthens the Liang-Kleeman heuristic balance
reasoning by showing its formal analogy with the Horowitz-Esposito
thermodynamic balance arguments. We showed that Eqs. (\ref{de}),
(\ref{IMPO1}), and (\ref{dif22}) within the Liang-Kleeman approach correspond
to Eqs. (\ref{de2}), (\ref{IMPO2}), and (\ref{dif1}) within the
Horowitz-Esposito approach, respectively. In particular, we have conveniently
recast the Liang-Kleeman formalism in Eq. (\ref{IMPO1}),%
\begin{equation}
\frac{dH_{1}}{dt}+\frac{dH_{2}}{dt}=\frac{dH}{dt}+\left[  T_{1\rightarrow
2}+T_{2\rightarrow1}\right]  \text{,} \label{B}%
\end{equation}
and underlined that the sum $T_{1\rightarrow2}+T_{2\rightarrow1}$ is a flow
$d_{t}I$ in Eq. (\ref{dif22}) that adds up additively to the time derivative
of mutual information as given in Eq. (\ref{MI}). Finally, we have compared
the thermodynamical interpretation of Schreiber's transfer entropy as provided
by Prokopenko \textit{et} \textit{al}. in \cite{prokopenko2013} to both the
Liang-Kleeman and the Horowitz-Esposito works. We have conveniently recast the
Prokopenko \textit{et} \textit{al}. formalism in Eq. (\ref{IMPO3}),%
\begin{equation}
\sigma^{X}+\sigma^{Y}=\Delta S+\left[  t_{Y\rightarrow X}\left(  n+1\right)
+t_{Y\rightarrow X}\left(  n+1\right)  \right]  \text{.} \label{C}%
\end{equation}
We provided here a quantitative reexamination leading to the essential
thermodynamical similarities among the Liang-Kleeman \cite{liang2005}, the
Horowitz-Esposito \cite{horowitz2014}, and the Schreiber \cite{schreiber2000}
works. Analogies are summarized in Eqs. (\ref{A}), (\ref{B}), and (\ref{C}).

In synthesis, the main results of our scientific effort can be outlined as follows:

\begin{itemize}
\item We established a novel link between the Liang-Kleeman analysis
\ \cite{liang2005} and the Horowitz-Esposito analysis \cite{horowitz2014}
concerning entropic balance arguments in the study of information transfer in
complex dynamical systems.

\item We strengthened the heuristic Liang-Kleeman information flow analysis
\cite{liang2005} with the support of fundamental thermodynamical arguments
borrowed from the Horowitz-Esposito work \cite{horowitz2014}.

\item We deepened our comprehension of the connection between the (newly
thermodynamically grounded) Liang-Kleeman information flow \cite{liang2005}
and the thermodynamical analog of Schreiber's transfer entropy
\cite{schreiber2000} as presented by Propokenko \textit{et} al. in
\cite{prokopenko2013}.
\end{itemize}

In conclusion, we think that our effort to emphasize such a unifying
theoretical picture of thermodynamical nature more or less explicit in
different works is extremely important for describing and, most of all,
understanding information flows between components of complex systems of
arbitrary nature from a foundational point of view.

\begin{acknowledgments}
C. Cafaro thanks Erik Bollt and Jie Sun for helpful discussions concerning
causality inference problems. Finally, constructive criticisms from two
anonymous Referees leading to an improved version of this manuscript are
sincerely acknowledged by the Authors.
\end{acknowledgments}

\bigskip

\bigskip

\end{document}